# Colloidal Joints with Designed Motion Range and Tunable Joint Flexibility

Indrani Chakraborty, Vera Meester, Casper van der Wel, and Daniela J. Kraft*

*Soft Matter Physics, Huygens-Kamerlingh Onnes Laboratory, Leiden Institute of Physics, PO Box 9504, 2300 RA Leiden, The Netherlands.*

*Kraft@Physics.LeidenUniv.nl*

**Abstract:** The miniaturization of machines towards the micron and nanoscales requires the development of joint-like elements that enable and constrain motion. We present a facile method to create colloidal joints, that is, anisotropic colloidal particles containing surface mobile DNA linkers that control the motion range of bonded particles. We demonstrate quantitatively that we can control the flexibility of these colloidal joints by tuning the DNA linker concentration in the bond area. We show that the colloidal shape controls the range of motion that these colloidal joints enable, due to the finite-sized patch of bonded linkers that cannot cross regions of high curvature. Finally we demonstrate the potential of the colloidal joints for programmable bottom-up self-assembly by creating flexible colloidal molecules and colloidal polymers. The reconfigurability and motion constraint offered by our colloidal joints make them promising building blocks for the development of switchable materials and nanorobots.

**Introduction**

Macroscopic machines transform an input signal from an actuator into a defined motion pattern. They rely on stiff structural elements called links connected by components that allow and constrain their movement, known as joints. Depending on their design, joints enable angular (hinges), linear (sliders), or spherical motion. More complex 3D motion and thus functionality can be achieved by combining multiple types of joints. Replicating these pivotal objects on the nano and micrometer scale will not only facilitate the assembly of functional materials with novel phase behavior not observed for rigid building blocks,[1–6] but also be a crucial step for the development of nanorobotics, biosensors and drug delivery devices.[7–10]

On the nanometer scale, DNA origami has recently been used to fabricate joints, hinges, crank-sliders and Bennett linkages, which facilitated angular and linear motion, coupled rotational and linear motion, and a complex 3D motion pattern, respectively.[7,11] On the colloidal scale, however, only one mechanism of a lock-and-key type has been reported so far. This lock-and-key mechanism consists of particles with dimples, the "locks", that flexibly connect spheres, the "keys", by exploiting a shape specific attraction imposed by depletion interactions.[4,12–14] The number of locks per particle sets the branching of the final structure and enables the assembly of colloidal polymers[15] and porous liquid phases.[16] However, these colloidal ball bearings rely on the presence of a depletant, constrain the angular motion range of the "key" particles bonded to the "locks" and do not provide any specificity for assembly other than the shape recognition. Ideally, colloidal joints should feature strong and specific bonding to direct their position within a desired structure, while allowing flexibility and different types of motion constraints.

Here, we present an experimental realization of colloidal joints with controllable joint flexibilty by employing colloidal particles with surface mobile DNA linkers. We show that the particle shape can be used to control the range of motion, and thereby experimentally realize micron-sized ball joints, hinges, and plane sliders. This fascinatingly simple reconfiguration mechanism relies on the accumulation of DNA linkers in the bond area, which allows us not only to control the motion range of the colloidal joints but also their flexibility. We demonstrate the potential of our colloidal joints by assembling two reconfigurable structures: flexible colloidal molecules and colloidal polymers.

**Results and Discussion**

We realize colloidal joints by coating colloidal silica particles of various diameters (1.15 ± 0.05 µm, 2.06 ± 0.05 µm, 3.0 ± 0.25 µm) with a lipid bilayer consisting of DOPC and 10 mol % DOPE-PEG$_{2000}$ by addition of small unilamellar lipid vesicles (SUVs). We functionalize these particles by adding DNA strands with hydrophobic double cholesterol anchors which insert spontaneously into the lipid bilayer.[5] The other end of the hybridized DNA strands features an unpaired sticky end of 11 base pairs and projects outwards into solution, which enables strong and specific binding to particles functionalized with the complementary sequence. See Experimental section for details. At room temperature, the lipid bilayer is in the liquid phase since the transition temperature $T_m$ of DOPC is -17°C.[17] Therefore, both the lipids and the DNA linkers can diffuse freely in the lipid bilayer while being confined to the particle surface. The presence of DNA on the particle surface is visualized by fluorescence or confocal microscopy through fluorescent markers (6-FAM or Cy3) attached to the 3' end after the sticky sequence. This particle functionalization scheme, shown in Figures 1a and b, enables strong and specific binding by the DNA sticky ends while allowing motion of bonded particles with respect to each other – the crucial ingredients for colloidal joints.

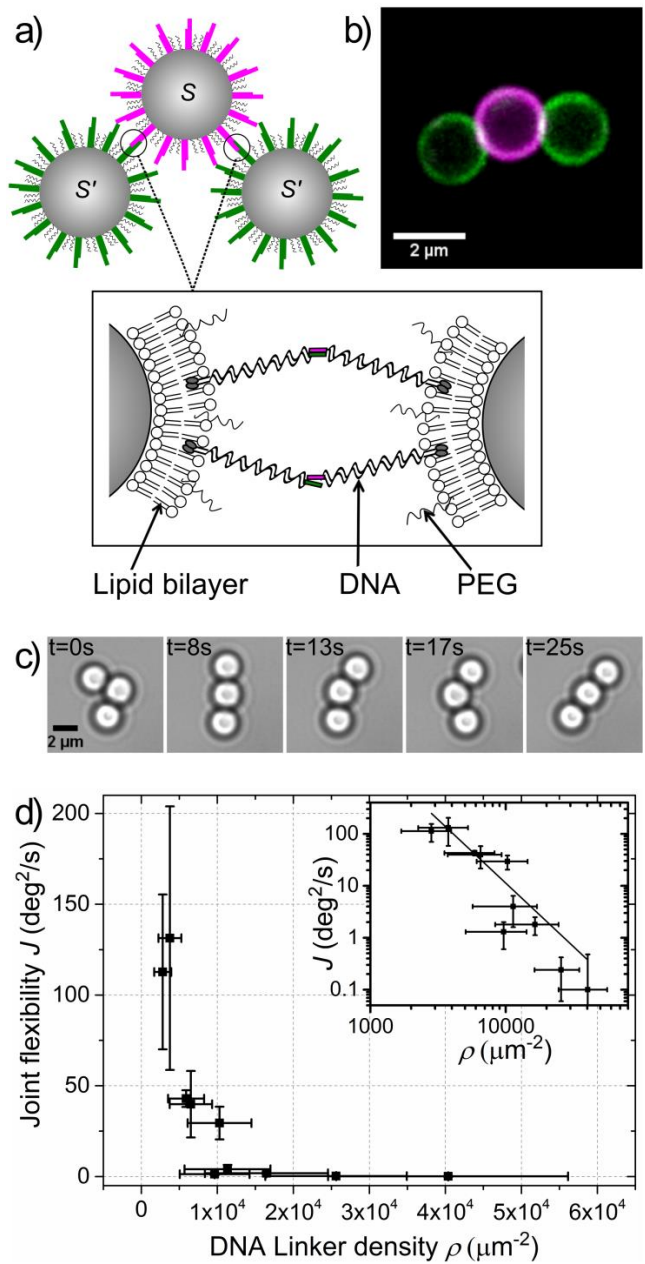

**Figure 1.** A spherical colloidal joint. a) Schematic diagram of the experimental particle configuration and surface functionalization. Green (*S'*) and magenta (*S*) show the fluorescence of the two complementary sticky ends. b) Confocal microscopy image of a colloidal joint with two bonded silica spheres functionalized with complementary DNA linkers. The white patches at the junctions indicate the bond areas. c) Time lapse sequence of a string of three particles showing

full angular motion. d) Joint flexibility $J$ decreases with increasing DNA linker density $\rho$ as $J \propto \rho^{-(2.4 \pm 0.5)}$. The inset shows the data in a log-log plot together with the least squares fit.

We first demonstrate that spherical colloids functionalized with mobile DNA linkers can act as flexible joints and measure their joint flexibility. To this end, we mix two batches of spheres functionalized with complementary DNA linkers $S$ and $S'$ and quantify the relative motion of the particles in a string of three spheres using particle tracking. Indeed, we find that two particles with linkers $S'$ bonded to a central sphere with complementary linkers $S$ can explore the full angular range with respect to each other, indicating that the central sphere acts as a spherical joint. See Figure 1c. Note that the motion of the here employed silica particles is essentially confined to the glass plane due to the small gravitational height of the particles.

We characterize the mobility or "joint flexibility" $J$ of such a colloidal joint by measuring the slope of the mean square displacement in the relative angle θ, $\langle \Delta \theta^2 \rangle / \Delta t$. The value of the joint flexibility of this colloidal joint is 200 deg$^2$/s, which relates to an effective diffusion constant of $D_{joint}$ = 0.03 μm$^2$/s through: $\boldsymbol{D_{joint} = d^2 J/8}$, where $d$ is the particle diameter. For comparison, unbounded 2 μm silica particles in our experiments exhibit a diffusion constant of $D_{free}$ = 0.044 μm$^2$/s. This is lower than the expected diffusion constant from the Einstein-Smoluchowski equation, $D_{ES}$ = 0.24 μm$^2$/s, possibly due to friction with the Pluronic F-127 polymer coating on the glass surface required to prevent particle adhesion (see Experimental section). The bonding with the colloidal joint slows down the thermal motion of the particles even more, likely due to viscous forces that hinder the motion of the linkers through the lipid membrane. A larger number of linkers participating in the bond should then decrease the diffusion constant more strongly and may provide a handle to tune the joint flexibility.

To test this idea, we study the effect of the linker surface density $\rho$ on the joint flexibility $J$ by preparing colloidal samples with different linker densities. We estimate the linker density by comparison of the fluorescent signal emitted by the DNA linkers of a given sample with the signal of particles with a known quantity of fluorescent DNA linkers. See Experimental section for details. This allows us to investigate the dependence of the joint flexibility on the linker density as shown in Figure 1d. All linker densities were measured in the absence of bonding, since bonding quickly leads to an increase of the concentration of the linkers in the bond area as indicated by the appearance of a bright fluorescent patch, see Figure 1b. We therefore measured the bond flexibility 1 hour after mixing of particles with complementary linker functionalization. An alternative way to restrict the linker density at the bond area is to bond particles with surface bound DNA linkers to colloidal joints.

Our first observation is that the spread in the linker density within a sample is rather broad, likely due to the random distribution of cholesterol anchors over the particles in a sample. At very low linker densities below $10^3$ linkers/$\mu m^2$, the probability for bonding between particles with complementary linkers drops significantly and we do not observe bond formation within 3 hours. At linker densities above $(3.8 \pm 1.5) \times 10^3$ linkers/$\mu m^2$ we find that the joints gradually stiffen up with increasing linker density. A least square fit of the experimental data with a power law function yields $J \propto \rho^{-(2.4 \pm 0.5)}$. Above a critical linker density of $(1.1 \pm 0.6) \times 10^4$ linkers/$\mu m^2$ the motion of the particles essentially arrests and the joint functionality is lost. Thus, the optimal linker density in which binding occurs sufficiently rapidly while providing adequate bond flexibility lies roughly between $10^3$-$10^4$ linkers/$\mu m^2$, at least for our 2 $\mu$m silica particles. We observed the threshold values to depend on the particle size and surface properties, but more systematic results are required to establish a conclusive trend.

Various effects may cause this loss of mobility at high linker densities: firstly, a larger number of bonded linkers increases the viscous drag of the bond patch through the lipid bilayer and therefore hinders the diffusion of the particles relative to each other. While the maximum patch area $A_p$ within which complementary linkers can bond is only determined by geometric factors, $A_p = 2\pi rL$, where $r$ is particle radius and $L$ the DNA linker length, a higher linker density does enable the formation of more bonded DNA linkers and thus increased friction between bonded particles. Secondly, a higher linker concentration implies high concentrations of cholesterol anchors in the bond area. Cholesterol is known to decrease the lateral diffusivity in DOPC bilayers in the liquid disordered phase[18] and in the extreme case may locally induce a phase transition towards a liquid ordered phase, which would drastically inhibit the bond mobility.[19] By conducting similar experiments with DNA linkers having double stearyl ($C_{18}$) anchors instead of double cholesterol anchors, we could exclude that the increased cholesterol concentration is causing the decreased mobility. While the time scale for bonding was much larger than for particles functionalized with cholesterol anchors, we again observed that high linker densities led to an increase in the joint stiffness and the formation of fractal aggregates, while low linker densities ensured mobility (see supporting Figure S2). Finally, despite the net free energy gain from bonding two mobile linkers,[20] there is also a significant entropic cost from confining the linkers to the patch area, that we can estimate to be on the order of $13k_BT$.[21] A straightening of the membrane in the patch area by stretching and flattening of the membrane similar to what has been observed in giant unilamellar vesicles[21] would increase the translational and combinatorial entropy[20,21] of the bonded linkers while simultaneously increasing the friction between the membrane in the centre of the bond area and the particle surface. Entropic forces may thus impact the joint flexibility as well.

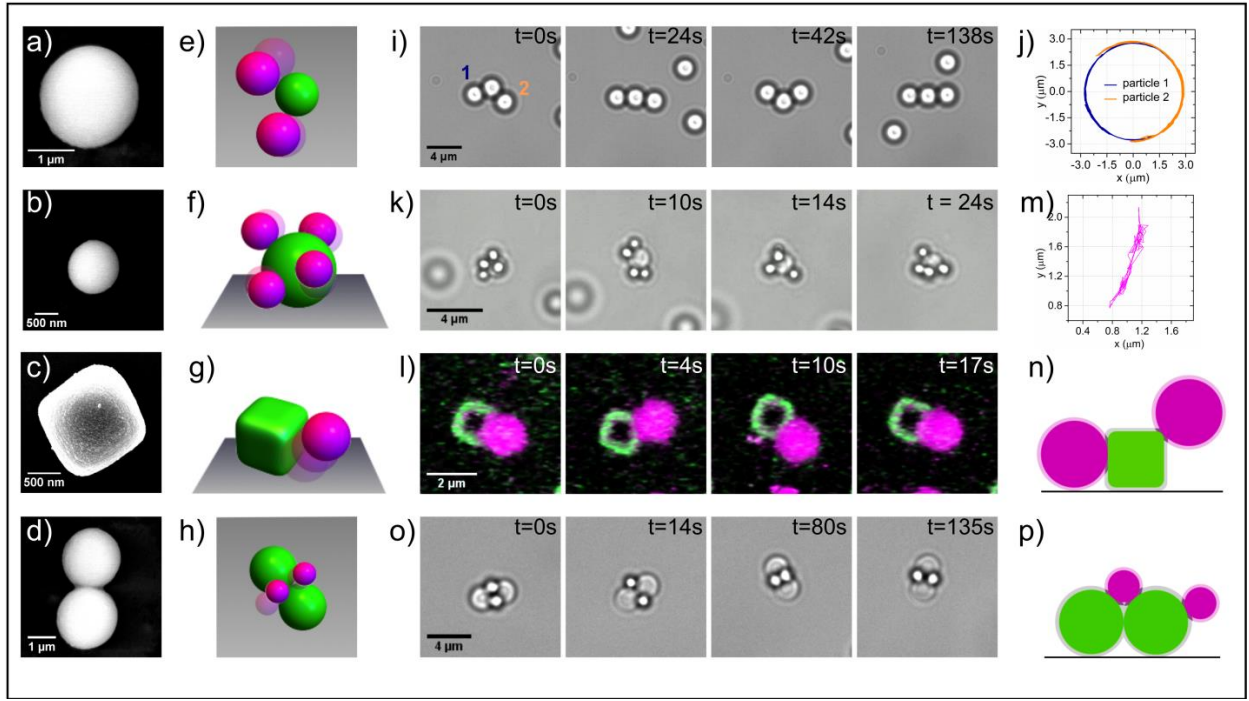

**Figure 2.** The shape of the colloidal joint determines the range of motion. a)-d) Scanning electron micrograph (SEM) of a 2 μm silica sphere, 1 μm silica particle, hollow 1 μm silica cube and dumbbell composed of 2 μm silica particles respectively. e)-h) Schematic diagrams of a spherical colloidal joint, a colloidal ball joint, a slider joint and a dumbbell hinge joint respectively. i) Still images from Supporting movie S1 showing a colloidal joint with 2π radians motion in 2D composed of 2 μm silica particles with mobile linkers. j) The trajectory of the two bonded particles in i) with respect to the central colloidal joint. k) Time lapse of a colloidal ball joint exhibiting 4π steradian motion in 3D composed of a 2 μm silica particles with mobile linkers and 1 μm polystyrene with surface bound DNA linkers (supporting movie S2). l) A slider type joint between a 1 μm silica sphere and a hollow 1 μm silica cube both with mobile DNA linkers (supporting movie S3). m) The trajectory of the silica particle (magenta) in l) shows motion only on one face of the cube. n) Schematic diagram showing that the bond patch area for a sphere at the corner of the cube is much more compact than that for a sphere on one face of the cube. This makes crossing of the corners entropically unfavourable. o) A colloidal hinge formed from a silica dumbbell of 2 μm spheres restricts the motion of two 1 μm polystyrene particles along the neck of the dumbbell (supporting movie S4). p) Schematic showing the double bond

areas for a sphere constrained to move along the neck of the dumbbell while a sphere on the positive curvature side has only one compact single bond area.

The formation of a patch of bonded linkers allows us furthermore to exploit the particle shape of the colloidal joint to constrain the range of motion. By choosing the appropriate shape (Figure 2a-d), different types of joints, similar to the ones available in macroscopic machines, could be created (Figure 2e-h). As shown above, a spherical colloidal joint allows the full $2\pi$ radians motion of bonded particles that move in quasi 2D due to their low gravitational height (See Figure 2i-j and supporting movie S1). Bonding lighter polystyrene spheres 1 µm in diameter lifts this constraint and we can observe the full $4\pi$ steradians range of motion in 3D, see Figure 2k and supporting movie S2. A spherical particle with surface mobile linkers therefore acts like a perfect ball and socket or spherical joint. Unlike a mechanical or colloidal ball and socket joint[12] where the ball fits into the groove of the socket for multidirectional movement, these colloidal analogues allow unrestricted continuous movement over the entire surface.

In contrast, colloidal joints with regions of different curvature (positive, flat or negative) can restrict the motion to areas with a sufficiently lower curvature. Once a particle has formed multiple links with a colloidal joint, the linkers will be distributed in a patch area that is set by the particles' geometry and the maximum distance over which two linkers can still be tethered (see supporting information). To diffuse into a region of higher positive curvature, this bond area needs to adapt to the new geometric situation by compacting. However, a compaction to half of the patch radius, for example, would require an entropy increase per linker of 1.4 $k_BT$ neglecting any crowding or multi-linker effects for 2 µm spherical silica particles and DNA linker length 20 nm.[20,21] Diffusion into regions of lower curvature is permitted, but once the bond area has extended, a return to regions of significantly higher curvature becomes unlikely. Thus, the patch of linkers guides and confines the bonded particle to regions of lowest curvature on the colloidal joint and allows us to constrain the range of motion.

We demonstrate this effect by utilizing cubic silica colloids with round edges fixed to a glass cover-slip and functionalized with strand *S'*, and dispersed silica spheres with strand *S*. After binding to the cubic colloid, the spherical particle freely diffuses on one side of the cube as revealed by particle tracking, but it does not move to another side, see Figure 2l-m and supporting movie S3. While the flat side of the cube allows motion without any restrictions, diffusion over the more strongly curved edges or corners is prohibited due to the significantly smaller patch size in these regions, see Figure 2n. The cubic particles therefore act as colloidal sliders. Similarly, colloidal dumbbells confine bonded spheres to the negatively curved area in the centre due to the two patches on either lobe of the dumbbell. A colloid with a dumbbell shape thereby only allows angular motion around its waist, essentially acting as a colloidal hinge (Figure 2o-p and supporting movie S4).

Finally, we demonstrate the potential of these colloidal joints for creating reconfigurable structures by assembling two exciting objects: flexible colloidal molecules and colloidal polymers. To obtain colloidal polymers, we mix 2 μm silica particles with mobile DNA linkers and 1 μm polystyrene particles with surface-bound DNA linkers (linker density $884 \pm 248$ μm$^{-2}$) in a ratio of 1:3. This size and number ratio produced a mixture of chains and clusters. The colloidal polymer, shown in Figure 3a and supporting movie S5, is a floppy chain which can fully reconfigure and even fold onto itself. The very low persistence length of our colloidal polymers is in contrast to previous realizations of flexible 1 D structures from gold nanorods[22] and dimpled spheres.[12] This opens up the possibility to use these linear chains in a hierarchical folding scheme. Encoding additional information in the form of multiple types of DNA linkers into the sequence of the individual joint elements will allow subsequent assembly and the formation of complex 3D architectures analogous to protein folding.[23] The yield of colloidal polymers could be increased by choosing a size ratio that geometrically restricts bonding of only two larger particles to one smaller one, or by using external fields to align the particles before linking.[24]

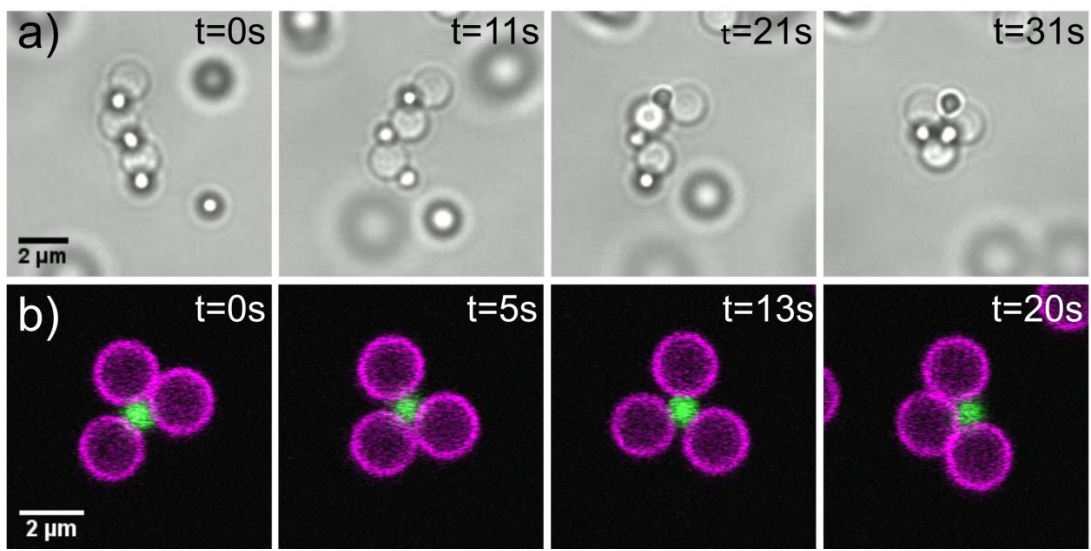

**Figure 3.** Self-assembly of reconfigurable structures. a) A floppy colloidal polymer consisting of 2 μm silica spheres with mobile DNA linkers and 1 μm polystyrene with surface bound DNA linkers reconfigures and folds onto itself. (See also movie S5) b) A flexible colloidal molecule composed of 2 μm (magenta) and 1 μm (green) silica spheres with mobile DNA linkers. (See also movie S6.)

At more extreme number ratios, for example by mixing 1 μm and 2 μm silica with complementary mobile DNA linkers in a ratio of 1:8, the formation of clusters is favoured.[25] Due to the colloidal joints at the centres, these colloidal clusters exhibit mobility of the outer spheres, essentially forming flexible colloidal molecules, see Figure 3b and supporting movie S6. By tuning the size ratio, various types of colloidal molecules may be assembled[26] and used as flexible building blocks to study novel crystal structures[1] and phase behaviour.[3] These simple yet exciting objects are just one example of the wide variety of reconfigurable structures that come within reach through the use of colloidal joints.

**Conclusions**

In summary, flexible colloidal joints were produced by embedding DNA linkers in lipid bilayers encapsulating micrometer-sized silica particles. The formation of a dense patch of linkers in the bond area leads to control over the flexibility of the joint and the range of motion for different particle shapes. We quantified the joint flexibility using particle tracking and observed that the joint flexbility depends on the linker density in the bond area. Controlling the maximum linker density by using colloids with surface bound linkers in combination with the colloidal joints thereby allows selection of the bond flexiblity. The finite patch size furthermore allows the restriction of motion on the particle surface by implementing joints with different local curvatures. Regions that allow expansion of the patch area act as sinks for the particles, thereby enabling the realization of colloidal joints, hinges and sliders. We demonstrated the potential of colloidal joints by assembling flexible colloidal polymers and colloidal molecules. The controlled joint stiffness and motion range provided by the colloidal joints makes them exciting elements for the development of reconfigurable materials and nanorobots.

**Experimetal Section**

**Materials.** The lipids 1,2 dioleoyl-sn-glycero-3-phosphocholine (DOPC), 1,2-dioleoyl-sn-glycero-3-phosphoethanolamine-N-[methoxy(polyethylene glycol)-2000](ammonium salt) (DOPE-PEG2000), 1,2-dioleoyl-sn-glycero-3-phosphoethanolamine-N-(lissamine rhodamine B sulfonyl) (ammonium salt) (DOPE-Rhodamine), 1,2-dioleoyl-sn-glycero-3-phosphoethanolamine-N-(carboxyfluorescein) (ammonium salt) (DOPE-Fluorescein) and Ganglioside GM1 (Ovine Brain) were obtained at >99% purity from Avanti Polar Lipids, Inc. The different DNA strands used (Eurogentec) had the following sequences: a) the A strand: Cholesterol-TEG-5'-TTT-ATC-GCT-ACC-CTT-CGC-ACA-GTC-AAT-CTA-GAG-AGC-CCT-GCC-TTA-CGA-*GTA-GAA-GTA-GG*-3'-6FAM b) The B strand: Cholesterol-TEG-5'-TTT-ATC-GCT-ACC-CTT-CGC-ACA-GTC-AAT-CTA-GAG-AGC-CCT-GCC-TTA-CGA-*CCT-ACT-TCT-AC*-3'-Cy3 and c) the C strand: Cholesterol-TEG-3'-TTT-TAG-CGA-TGG-GAA-GCG-TGT-CAG-TTA-GAT-CTC-TCG-GGA-CGG-AAT-GC-5'. The A and B strands have 11 base pair long complementary sticky ends at the 3' ends (denoted by italic characters) and are labelled with a green fluorescent dye 6-FAM, i.e. 6-Carboxyfluorescein (excitation: 488 nm, emission: 521 nm) and the red Cy3 dye (excitation: 561 nm, emission: 570 nm), respectively. For the particles with surface bound DNA linkers, the two DNA strands used were a) the D strand: 5'-CGT-AAG-GCA-GGG-CTC-TCT-AGA-TTG-ACT-GTG-CGA-AGG-GTA-GCG-ATT-TT-3'-TEG-Biotin and b) the E strand: 5'-Cy3-TTT-ATC-GCT-ACC-CTT-CGC-ACA-GTC-AAT-CTA-GAG-AGC-CCT-GCC-TTA-CGA-*CCT-ACT-TCT-AC*-3'.

**Particle synthesis.** Silica spheres of 1.15 ± 0.05 μm diameter (with Fluorescein if required) were synthesized using a modified Stöber method.[27] Commercial silica spheres were purchased from Microparticles GmbH (2.06 ± 0.05 μm, 3.0 ± 0.25 μm). Hollow silica cubes with rounded edges and an edge length of 1.24 ± 0.08 μm were synthesized by coating haematite cubes made by a sol-gel method[28] with silica[29] and dissolving the haematite cores in HCl.[29] Silica dumbbells were obtained from aggregated 2μm silica colloids. Polystyrene spheres of 1.05 ± 0.02 μm diameter containing carboxylic acid surface groups were synthesized following a surfactant free radical polymerization protocol.[30] The spheres were coated with neutravidin by adding 8 µL of 5 g/L neutravidin and 10 µL of freshly dissolved 1 g/L EDC to a suspension of 1 mg particles in 500 µL MES buffer (10mM MES, 10mM NaCl, 0.5 wt.% Pluronic F-127, pH = 4.7). The coupling reaction was allowed to proceed on the vortex mixer for 120 min. at room temperature and washed three times.

**Functionalization of colloidal particles with mobile linkers.** For functionalization with surface mobile DNA linkers, we adapted a protocol from reference[5]. For the spherical colloidal joints, small unilamellar lipid vesicles (SUVs) were prepared from mixtures of DOPC and DOPE-PEG$_{2000}$ in a 90:10 molar ratio in chloroform. If necessary, 0.001% mole fraction of DOPE-Rhodamine or DOPE-Fluorescein was added to increase fluorescence. After desiccation in vacuum, the lipids were re-suspended in HEPES buffer (10 mM HEPES, 47 mM NaCl, 3 mM NaN3, pH = 7.01) for 30 mins to obtain a 3g/L solution. Then the lipid mixture was extruded 21 times through two stacked polycarbonate filters (Whatman) with 30 nm pore size to obtain SUVs. For the functionalization of the hollow silica cubes, a composition of 58.6% DOPC, 6.9% DOPE-PEG$_{2000}$ and 34.5% GM1 (by mole-fraction) was used. We hybridized 50 µM solutions of the A or B DNA strand with the C strand in a 1:1.5 volume ratio by heating the solution to 90°C and cooling down at 1°C/min. The hybridized strands consisted of a 47 base pair long double-stranded central part, with double cholesterol anchors connected through TEG (tetraethylene glycol) spacers at one end and an 11 base-pair long single-stranded sticky part at the other end which could bind to a complementary sticky end. *S'* strand is obtained by hybridizing A and C and *S* strand by hybridizing B and C. For particles with surface bound linkers, the E strand and the biotinylated D strand were hybridized separately under similar conditions.

The SUVs were added to an equal volume of 5g/L colloidal silica solution and put on a rotating tumbler for 40 mins at 9 rotations/min to prevent sedimentation by gravity. After that the particles were centrifuged for 5 min at 494 rcf and washed with HEPES. Required amounts of DNA were added to the SUV encapsulated particles and the resulting solution was again kept on the tumbler for 1 hour. Finally the particles were washed three times in HEPES by centrifugation to remove excess dye and lipids.

**Functionalization of colloidal particles with bound linkers.** For producing the colloids with surface bound linkers, 50 µl of 5 g/L neutravidin coated polystyrene particles were incubated with appropriate amounts of 2.08 µM pre-hybridized biotinylated DNA in PBS buffer. The sample was incubated at 55 °C for 30 minutes, cooled to room temperature and washed 3 times.

**Functionalization of hollow silica cubes with mobile linkers**. The hollow silica cubes were at first injected onto a cleaned glass cover slip and the PVP layer on the cubes was decomposed at 500° C for 30 mins. The cover-slip containing the dried cubes was introduced into a microscope sample holder and 600 µl of HEPES was added. 150 µl SUVs composed of DOPC, DOPE-PEG$_{2000}$ and GM1 were mixed with 2 µl of Choleratoxin Alexa Fluor 488 dye and 10 µl of 4.34 µM hybridized DNA with 6-FAM and incubated for 1 day. The SUVs were then added to the cubes and the final solution in the sample holder was heated for 30 mins.

**Imaging and tracking.** Particles were imaged in sample holders having a hydrophobized and passivated glass as the bottom surface. Hydrophobization of the glass surface was done using Surfasil (a siliconizing agent), followed with passivation by adding 5% w/v Pluronic F-127 to the holder, storing it for 30 mins and rinsing with water. Imaging was done using an inverted Nikon TI-E A1 confocal microscope with 8kHz resonant scanner and 100x objective lens (NA 1.4).

Particle tracking was done using Trackpy,[31] an open source Python based implementation of the Crocker and Grier algorithm.[32]

**Measurement of linker density.** Linker densities on the colloidal particles were measured in a comparative way: first, we estimated the linker density of a highly fluorescent reference sample. To this end, we quantified the amount of fluorescent DNA linkers left in the supernatant after particle functionalization (collected from three washing cycles) using a fluorimeter calibrated with a known concentration of DNA linkers. The DNA linker density on the reference colloids was obtained by subtracting the DNA left in the supernatant from the amount of added DNA. We then obtained the linker density of all other colloids by comparison of their fluorescence intensity collected in the photomultiplier tubes of the confocal microscope to this reference sample, using identical conditions of laser power and gain in the linear response regime.

For the polystyrene particles with surface bound linkers, we measured the amount of neutravidin on the polystyrene particles to be $\sim 10^5$ molecules/particle following Kada *et al.*[33] Assuming that all the biotinylated DNA linkers are bound to neutravidin molecules at a low DNA density ($\sim 10^3$ linkers/$\mu m^2$), we can approximate the number of surface bound linkers per polystyrene particle to be equal to the number of added linkers per particle. By comparing the fluorescence intensity of samples with higher linker densities with this reference sample using confocal microscopy as outlined above, we estimated the approximate linker density on all other samples.

**Supporting Information:**

Figures S1 and S2 and Movies S1-S6.

**Acknowledgement:**

This work was financially supported by the Netherlands Organisation for Scientific Research (NWO) through VENI grant 680-47-431 and as part of Frontiers of Nanoscience program (NWO/OCW), and through the Sectorplan for Physics and Chemistry. We thank Stef van der Meulen and Mirjam Leunissen for introducing us to the functionalization protocol published in reference[5] and Melissa Rinaldin for helping with the functionalization of the cubic colloids.

# Colloidal joints with Designed Motion Range and Tunable Joint Flexibility

Indrani Chakraborty, Vera Meester, Casper van der Wel and Daniela J. Kraft*

*Soft Matter Physics, Huygens-Kamerlingh Onnes Laboratory, Leiden Institute of Physics, PO Box 9504, 2300 RA Leiden, The Netherlands.*

## Calculation of bond area of colloidal joints

We consider bonding of two spheres of radii $r_1$ and $r_2$ by mobile DNA linkers. We presume that the maximum distance $d_{max}$ at which two linkers still bond is twice their length $l$ ($l$ = 20 nm). This parameter contains small corrections to the actual linker lengths to take the overlap at the sticky end and insertion into the lipid membrane into account. The maximum bond area defines a spherical cap on each sphere within which linkers can bind to their complementary linkers on the other sphere. To calculate the bond area of the colloidal joints, we express the height $h$ in terms of the cap radius $a$ and the radius of the sphere $r$.

$$h(r, a) = r - \sqrt{r^2 - a^2} \tag{1}$$

Setting $h_1 + h_2 = 2l$, we obtain,

$$a(r_1, r_2) = 2 \frac{\sqrt{l(r_1 - l)(r_2 - l)(r_1 + r_2 - l)}}{(r_1 + r_2 - 2l)} \tag{2}$$

For equal sized spheres, $r_1 = r_2 = r$, equation 2 simplifies to:

$$a(r) = \sqrt{2lr - l^2}$$

The bond areas are therefore given by:

$$A_b(r_1) = 2\pi r_1 \cdot h_1(r_1, a) = 2\pi r_1 \left( r_1 - \sqrt{r_1^2 - \frac{4l(r_1-l)(r_2-l)(r_1+r_2-l)}{(r_1+r_2-2l)^2}} \right)$$

$$A_b(r_2) = 2\pi r_2 \cdot h_2(r_2, a) = 2\pi r_2 \left( r_2 - \sqrt{r_2^2 - \frac{4l(r_2-l)(r_1-l)(r_1+r_2-l)}{(r_1+r_2-2l)^2}} \right) \tag{3}$$

For equal sized spheres, this reduces to:

$$A_b = 2\pi r l \tag{4}$$

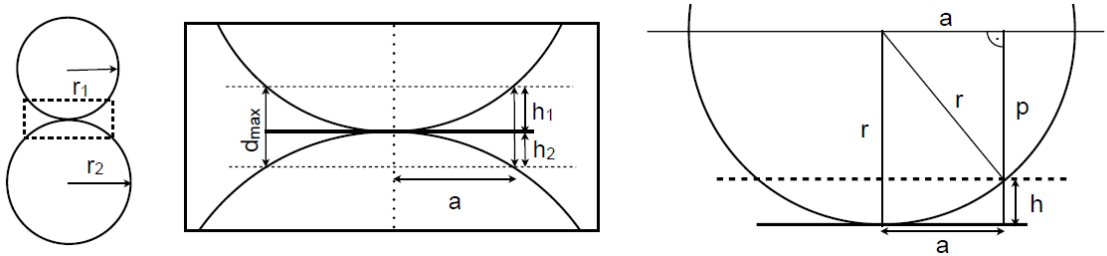

**Figure S1**. Geometry of two binding spheres with DNA linkers.

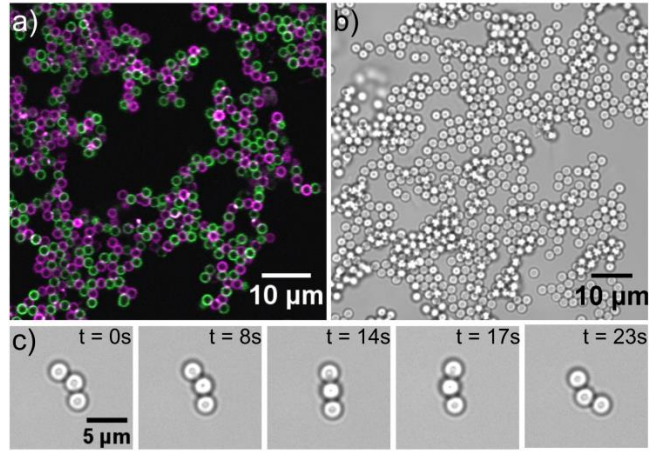

**Figure S2.** Colloidal joints with two different types of DNA anchors. a) Joints with a very high density of mobile DNA linkers ($>10^4$ $\mu m^{-2}$) containing double cholesterol anchors are immobile and form fractal aggregates at high linker density. b) Similar behaviour is shown by joints with DNA containing double stearyl ($C_{18}$) anchors. c) At low DNA linker densities (~ $10^3$ $\mu m^{-2}$) colloidal joints with stearyl anchored DNA show reconfigurability similar to joints containing cholesterol anchored DNA.

**Video Captions**

**S1:** A three particle colloidal string composed of colloidal joints, which consist of 2 μm silica particles functionalized with complementary DNA linkers. The video has been sped up 3.7 x times for viewing convenience.

**S2:** A colloidal ball joint composed of a 2 μm silica particle with mobile DNA linkers and 1 μm polystyrene particles with bound DNA linkers as joint elements.

**S3:** A colloidal slider joint composed of a 1 μm silica particle and a hollow 1 μm silica cube. Both have mobile DNA linkers. The motion of the spherical silica particle is restricted to sliding along one face of the cube.

**S4:** A colloidal hinge composed of two polystyrene particles with surface bound DNA linkers undergoing constrained motion along the neck of a silica dumbbell.

**S5:** A reconfigurable colloidal polymer composed of 2 μm silica with mobile DNA and 1 μm polystyrene with surface-bound DNA.

**S6:** A colloidal molecule composed of 2 μm and 1 μm silica particles with mobile DNA linkers.